\documentclass[12pt]{iopart}

\usepackage{graphicx}
\usepackage{comment}
\usepackage{iopams}  

\newcommand{\bra}[1]{\langle #1 |}
\newcommand{\ket}[1]{| #1 \rangle}
\newcommand{\braket}[2]{\langle #1 | #2 \rangle}

\newcommand{\hrho}{\hat{\rho}}

\newcommand{\hA}{\hat{A}}
\newcommand{\hM}{\hat{M}}

\newcommand{\hrec}{\tilde{W} (\cdot)}

\newcommand{\ha}{\hat{a}}
\newcommand{\hn}{\hat{n}}
\newcommand{\hN}{\hat{N}}

\newcommand{\combi}[2]{\left( \begin{array}{c} #1 \\  #2  \end{array} \right)  }

\newcommand{\dw}{d\tilde{W}}

\newcommand{\kptt}{\ket{\tilde{\psi} (t) }}

\newcommand{\mean}[1]{\langle  #1  \rangle  }

\begin{document}

\title[Simultaneous continuous measurement]{Simultaneous continuous measurement of photon-counting and homodyne detection on a free photon field: dynamics of state reduction and mutual influence of measurement backaction}

\author{Yui Kuramochi$^1$, Yu Watanabe$^2$, and Masahito Ueda$^3$}
\address{$^1$Department of Physics, University of Tokyo, 7-3-1 Hongo, Bunkyo-ku, Tokyo 113-0033, Japan}
\address{$^2$Yukawa Institute for Theoretical Physics, Kyoto University, Kitashirakawaoiwake-cho, Sakyo-ku, Kyoto 606-8502, Japan}
\ead{kuramochi@cat.phys.s.u-tokyo.ac.jp}

\begin{abstract}
We analyze a simultaneous continuous measurement of photon-counting and homodyne detection.
The stochastic master equation or stochastic Schr\"odinger equation describing the measurement process includes
both jump-type and diffusive-type stochastic increments.
Analytic expressions of the wave function conditioned on homodyne and photon-counting records are obtained,
yielding the probability density distributions and generating functions of the measurement records.
Formula for the expectation values of the homodyne records conditioned on a photon-counting event is also derived
which quantitatively describes the measurement backaction of photon-counting on the homodyne output.
The obtained results are applied to typical initial states --- coherent, number, thermal, and squeezed states.
Monte Carlo simulations of the measurement processes are also presented 
to demonstrate the dynamics of the combined measurement process.
\end{abstract}
\submitto{\JPA}

\section{Introduction}

Wave-particle duality is the hallmark in the quantum theory of radiation.
In real experiments,
wave and particle properties of radiation can be observed by homodyne and photon-counting
measurements, respectively.
In this paper, we investigate this wave-particle duality \textit{during the measurement process} 
by analyzing a model of simultaneous photon-counting and homodyne measurement.
The primary motivation of this work is to investigate the dynamics of state reduction 
during such simultaneous measurements 
and to clarify the mutual influence of one type of measurement on the outcomes of the other.

From a measurement theoretical viewpoint,
the analysis of such a simultaneous measurement process
should take into account the backaction of the measurement 
and continuity of the measurement with respect to time.
For the first point,
the classical theory of projection measurement~\cite{neumann}
is insufficient to describe the measurement 
and we need a more general framework such as defined in~\cite{davies_lewis}.
In addition, there are two output channels in this measurement 
which give rise to backaction of the system's wave function
and statistically correlate with each other through the backaction.
Such a correlation reflects the mutual influence of one type of measurement on the outcomes of the other.

As for the second point,
in the measurement process,
outputs are continuously recorded
and the system density operator is also continuously renewed 
due to the backaction of the measurement
depending on the outputs.
Such a measurement process is called a continuous quantum measurement.
The theory of the continuous quantum measurement has been studied from various standpoints.
(also see \cite{jacobs_steck1} for a review).
One is a mathematical 
approach~\cite{davies_qsp,diosi,wiseman_diosi,barchielli_paganoni_zucca}
in which general properties of a stochastic process of a state vector is studied.
There are also related works from the stand point of open quantum system.
A general relationship between quantum open systems
and quantum measurement~\cite{davies_book,kraus_wootters,zurek,presilla}
is known to hold
in the sense that non-selective evolution under a quantum measurement
can be identified with an open system's evolution.
In the continuous quantum measurement,
the non-selective time evolution corresponds to
a Lindblad master equation~\cite{lindblad,gorini_kossakowski_sudarshan}
which is a general form of a master equation
of an open quantum system which is coupled with a Markovian environment.
Monte Carlo wave-function approach~\cite{dalibard,gisin,breuer_petruccione,haroche_raimond}
utilizes this fact
for the numerical computations of the Lindblad master equation of an open system.
A path integral approach~\cite{mensky,chantasri} is also another important
formulation of the continuous measurement.

There are two distinct types of continuous measurement processes: 
jump-type and diffusive-type processes.
In the jump-type measurement, discontinuous state changes occur at discrete times.
A typical example is the photon-counting measurement~\cite{srinivas_davies,ueda,ueda_ogawa,imoto_ueda_ogawa},
where the coupling between the photon field and the detector is adjusted so that
the probability of more than one photon being detected during any infinitesimal
time interval (i.e., the resolution time)
is negligible.
Therefore, the photon-counting measurement consists of two fundamental processes:
no-count and one-count processes.
The state change in the no-count process is, however, different from that of measurement-free evolution due to the back action of the measurement.
The model of the photon-counting measurement can be also derived from
the cavity quantum electrodynamics setup~\cite{imoto_ueda_ogawa}
in which two-level atoms are successively driven into
an optical cavity and the level of the atoms is measured after
the interaction with the radiation in the cavity.
An experiment in the same spirit is performed by Haroche \textit{et al.}~\cite{haroche_qnd}.

On the other hand, in the diffusive-type measurement~\cite{wiseman_diosi,barchielli_book},
the state change is continuous.
Examples include
balanced homodyne measurement~\cite{carmichael1,wiseman_milburn1}
and continuous observations of the position of a particle~\cite{diosi}.
In classical stochastic processes,
the jump-type and diffusive-type continuous measurements correspond to
the Poissonian and Wiener processes, respectively~\cite{gardiner1,jacobs_book}

Mathematically, the hybrid type of continuous measurement with diffusive and jump outcomes
is also possible.
A general equation of the continuously observed system was
derived under general semigroup assumptions~\cite{barchielli_paganoni_zucca,barchielli_holevo}
and also from the continuous time limit of the discrete time process~\cite{pellegrini}.
The simultaneous measurement of photon-counting and homodyne detection analyzed in this paper
is an example of the hybrid type continuous measurement.
In quantum optical setup, this model is a generalization of the models of photon-counting~\cite{srinivas_davies,ueda,ueda_ogawa,imoto_ueda_ogawa}
and the homodyne detection~\cite{wiseman_milburn1}.
We note that a simultaneous measurement of photon-counting and homodyne detection
was discussed in \cite{carmichael_orozco,wiseman}
in a different context
and the simultaneous measurement of the two-level atom and the position coordinate is also discussed~\cite{viola_onofrio}.
We also note that continuous measurements driven by L\'evy processes 
are discussed in \cite{jacobs_steck2}

This paper is organized as follows.
In Sec.~2, we present the mathematical model of 
the simultaneous measurement of photon-counting and homodyne detection
and
derive an analytic expression of the conditional wave function.
In Sec.~3, we examine probability laws of measurement records 
by deriving the probability distributions and the generating functional for measurement outcomes.
In Sec.~4, 
we apply the obtained general expressions to typical initial quantum states, 
namely, coherent, number, thermal, and squeezed states.
We also present the results of Monte Carlo simulations for each of these initial states
to illustrate how the hybrid-type measurement backactions disturb the average photon number.
In Sec.~5, we summarize the main results of this paper.
In the Appendix, 
we show derivations of some formulas used in the main text.

\section{Simultaneous measurement of homodyne detection and photon-counting}
In this section, we consider a simultaneous measurement of photon-counting and homodyne detection
and derive the stochastic wave function conditioned on the measurement outcomes.
\subsection{Setup of the system}
\begin{figure}
\centering
\includegraphics[width=12cm,clip]{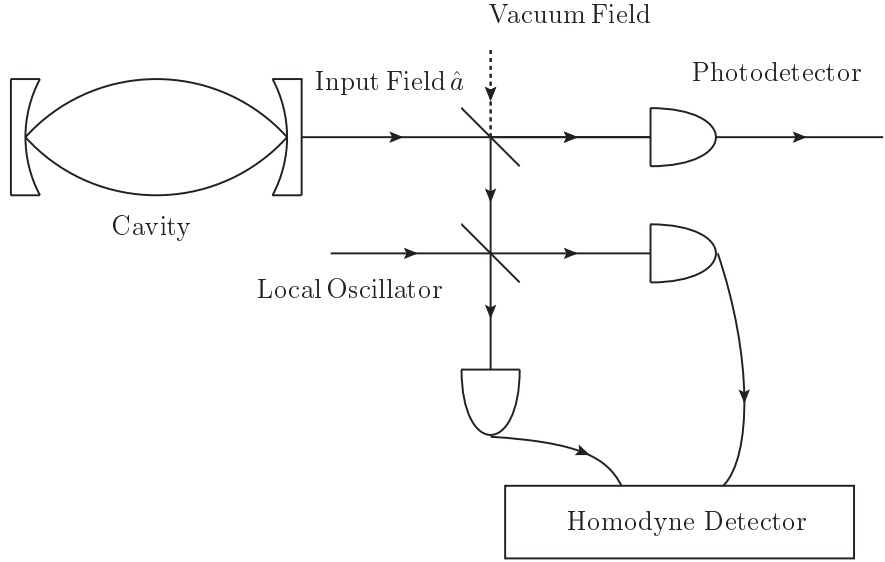}
\caption{Schematic illustration of the system.
	An input photon field enclosed in a cavity is simultaneously measured 
	by a photodetector and a balanced homodyne detector.
	See the text for details.
}
\label{chm1}
\end{figure}

The measurement scheme discussed in this section
consists of photon-counting and balanced homodyne detection,
as schematically illustrated in figure~\ref{chm1}.
A single-mode photon field
confined in a cavity,
described by the annihilation operator $\ha$, is divided by a beam splitter into two,
one of which is detected by a photodetector and the other is
superimposed by a local oscillator with amplitude $\beta$ and then
measured by a balanced homodyne detector.
We do not consider the pumping of the cavity field and
assume that the system is coupled only to the detectors.

The measurement process during an infinitesimal time interval $dt$
can be mathematically expressed by the following measurement operators:
\begin{eqnarray}
	\hM_{\dw} &=&  1 -\left( i \omega +  \frac{\Gamma}{2}  \right) \hn dt + \sqrt{\gamma_2}\ha \dw  , \label{chmmo1} \\
	\hM_1 &=& \sqrt{\gamma_1 dt} \ha  ,  \label{chmmo2}
\end{eqnarray}
where $\gamma_1>0$ and $\gamma_2>0$ denote 
the coupling strength of the photodetector and that of the homodyne detector, respectively,
and $\Gamma  := \gamma_1 + \gamma_2$ is the total loss rate of the cavity.
The unitary part of the time evolution is given by $ \omega \hn$, 
where $\omega$ is the detuning of the photon field with respect to the homodyne local oscillator $\beta$ and
$\hn := \ha^\dagger \ha$ is the number operator of the photon field.
The Planck constant $\hbar$ is set to be unity throughout this paper.
In equation~(\ref{chmmo1})
$\dw$ is the stochastic variable corresponding to the homodyne record
with the expression~\cite{jacobs_steck1,wiseman_diosi}
\begin{equation}
	\dw = \mean{\ha + \ha^\dagger} dt + dW ,
\end{equation}
where $\mean{\cdot}:=\mathrm{tr}[\hrho \cdot ]$
and $dW$ is the Wiener increment which obeys the It\^o rule $(dW)^2 = dt$~\cite{jacobs_steck2,gardiner_zoller,wiseman_milburn2}.
Equation~(\ref{chmmo1}) describes homodyne detection, 
while the measurement operator in equation~(\ref{chmmo2}) corresponds to the photodetection event.
Actually, the vacuum field $\hat{b}$ enters the mirror before the photodetector, 
but this effect can be neglected
because it does not contribute to the photodetector.

\subsection{Wave function for the no-count process}
Let the initial state vector of the system at $t=t_0$ be $\ket{\psi_0}$.
We consider the time evolution of the wave function under the condition
that the homodyne records are given by $\tilde{W} (\cdot)$
and that there is no photocount.
It is described by the measurement operator in equation~(\ref{chmmo1}).
Thus, the unnormalized wave function $\ket{\tilde{\psi} (t)}$ during the no-count process obeys
\begin{equation}
	\ket{ \tilde{\psi} (t+dt) } = \hM_{\dw} \kptt  
	= \left[ 1 - \left( i \omega +  \frac{\Gamma}{2}  \right) \hn dt + \sqrt{\gamma_2} \ha \dw (t)  \right] \kptt ,   \label{ncsse}
\end{equation}
where the tilde over $\psi$ indicates that the state vector is unnormalized.

By direct substitution, it can be shown that 
the solution of the no-count stochastic Schr\"odinger equation~(\ref{ncsse})
is given by 
\begin{eqnarray} \fl
	\ket{\tilde{\psi}(t)}&=&  e^{- \left( i \omega + \frac{\Gamma}{2}  \right)(t -t_0)\hn }
	\exp \left[ \sqrt{\gamma_2} \int_{t_0}^t e^{- \left( i \omega + \frac{\Gamma}{2}  \right)(t^\prime -t_0)} \dw (t^\prime) \ha
	-  \frac{\gamma_2}{2} \frac{1-e^{-( 2i \omega + \Gamma) (t- t_0) } }{ 2i \omega + \Gamma}   \ha^2  \right] \ket{\psi_0}  \nonumber \\
	\fl
	&=:& \hN (t, t_0; \hrec) \ket{\psi_0} . \label{defN}
\end{eqnarray}
Note that the $\hat{a}^2$ term in the exponential in equation~(\ref{defN})
is needed from the It\^o rule $\dw(t)^2 = dt$.

\subsection{Wave function for the $m$-count process}
We generalize the result obtained in the previous subsection 
to the conditional wave function under the $m$-count process.
Let us assume that the initial condition of the wave function is $\ket{\psi (t=0)} = \ket{\psi_0}$.
The stochastic Schr\"odinger equation for this general process is given by
\begin{equation} \fl
	\ket{\tilde{\psi} (t+dt) } = \left[ 1 - \left( i \omega +  \frac{\Gamma}{2}  \right) \hn dt + \sqrt{\gamma_2} \ha \dw  (t) \right] \kptt
	+ dN(t) \cdot( \sqrt{\gamma_2 dt} \ha -1)\ket{\tilde{\psi} (t)},
	\label{mcsse}
\end{equation}
where $dN(t)$ is defined by
\begin{equation}
	dN(t) := \left\{ \begin{array}{l}  1 \quad \mathrm{if \; a\; photocount\; occurs;} 
	\\ 
	0 \quad \mathrm{otherwise.} 
	\end{array}\right.
\end{equation}
This equation implies that for the no-count case $dN(t) = 0$ the wave function evolves in the same manner as
in~(\ref{ncsse}),
while for the one-count process $dN(t) = 1$ the wave function immediately after the photocount is
$\hM_1 \ket{\tilde{\psi}(t)}= \sqrt{\gamma_1 dt} \ha \ket{\tilde{\psi}(t)}.$
Thus the wave function under the condition that photocounts occur at times
$t_1, t_2 , \cdots , t_m$ during the time interval $(0,t)$
($0 < t_1< t_2 < \cdots < t_m <t$)
and that the homodyne records are $\hrec$ is given by
\begin{equation}
	\ket{\tilde{\psi}(t)} = (dt)^{m/2} \ket{ \tilde{\psi} (t;t_1,\, t_2 , \dots t_m ; \tilde{W}( \cdot ) )  } ,
\end{equation}
where
\begin{eqnarray} \fl
	\ket{ \tilde{\psi} (t;t_1,\, t_2 , \dots t_m ; \tilde{W}( \cdot ) )  }   \nonumber \\
	\fl =  \gamma_1^{m/2}\hN (t,t_m; \tilde{W}( \cdot ))  \ha  \hN (t_m,t_{m-1} ; \tilde{W}( \cdot )) \ha \cdots \ha \hN (t_2,t_1; \tilde{W}( \cdot ))  \ha \hN (t_1  ,0; \tilde{W}( \cdot )) \ket{\psi_0}.   \label{conpsi0}  
\end{eqnarray}
The product of the operators on the right-hand side (rhs) of equation (\ref{conpsi0}) can be simplified as
\begin{eqnarray}
	&\hN (t,t_m; \tilde{W}( \cdot ))  \ha  \hN (t_m,t_{m-1} ; \tilde{W}( \cdot )) \ha \cdots \ha \hN (t_2,t_1; \tilde{W}( \cdot ))  \ha \hN (t_1  ,0; \tilde{W}( \cdot ))  
	\nonumber \\ 
	&=     e^{   - \left(i \omega + \frac{\Gamma}{2}\right)  (    t_1  + t_2 + \cdots + t_m  ) } e^{ - \left(i \omega + \frac{\Gamma}{2}\right) t \hn }    \ha^m \exp \left[  A (t) \ha +   B(t) \ha^2 \right],   \label{conop1}  
\end{eqnarray}
where $A (t)$ and $B(t)$ are given by
\begin{eqnarray}
	A(t) &:=  \sqrt{\gamma_2} \int_0^{t} e^{ - \left(i \omega + \frac{\Gamma}{2}\right) t^\prime  } \dw (t^\prime ),  \label{defA} \\
	B(t) &:= - \frac{ \gamma_2 }{2} \frac{ 1- e^{ - ( 2i\omega + \Gamma) t}    }{2i\omega + \Gamma} .  \label{defB} 
\end{eqnarray}
Thus, the conditional $m$-count wave function is given by
\begin{eqnarray}
	\fl
	&\ket{ \tilde{\psi} (t;t_1,\, t_2 , \dots t_m ; \tilde{W}( \cdot ) )  } \nonumber \\ 
	\fl
	&= \gamma_1^{m/2} e^{   - \left(i \omega + \frac{\Gamma}{2}\right)  (    t_1  + t_2 + \cdots + t_m  ) } e^{ - \left(i \omega + \frac{\Gamma}{2}\right) t \hn }    \ha^m \exp \left[  A(t) \ha +   B(t) \ha^2 \right] \ket{\psi_0} 
	\nonumber \\
	\fl
	&= \gamma_1^{m/2} e^{   - \left(i \omega + \frac{\Gamma}{2}\right)  (    t_1  + t_2 + \cdots + t_m -mt  ) }
	\ha^m \exp \left[  \tilde{A} (t) \ha +   \tilde{B} (t) \ha^2 \right] e^{ - \left(i \omega + \frac{\Gamma}{2}\right) t \hn } \ket{\psi_0},
	\label{conpsi3}
\end{eqnarray}
where $\tilde{A}(t):= e^{ \left( i\omega + \frac{\Gamma}{2}  \right) t} A(t)$
and
$\tilde{B}(t):= e^{ \left( 2 i\omega +\Gamma  \right) t} B(t) $.
Note that the time dependence on photocounts of this wave function arises, 
aside from the c-number factor 
$e^{   - \left(i \omega + \frac{\Gamma}{2}\right)  (    t_1  + t_2 + \cdots + t_m  ) }$,
only through the number of photocounts $m$.

\section{Probability laws of measurement records}
\subsection{Probability density functions}
In this section, we will derive general results on the probability distributions of homodyne and photocount records.
From the general considerations on the measurement operators, the joint probability density of homodyne records $\hrec$ and photodetection times $ t_1 , t_2 , \cdots , t_m $ is given by
\begin{equation} \fl
	dt_1 dt_2 \cdots dt_m \times \mu_0 (\hrec) \times \braket{  \tilde{\psi} (t;t_1,\, t_2 , \dots t_m ; \tilde{W}( \cdot ) ) }{ \tilde{\psi} (t;t_1,\, t_2 , \dots t_m ; \tilde{W}( \cdot ) ) },
\end{equation}
where $\mu_0$ is the Wiener measure.
The square of the norm of the wave function is evaluated to be
\begin{eqnarray}\fl
	\braket{  \tilde{\psi} (t;t_1,\, t_2 , \dots t_m ; \tilde{W}( \cdot ) ) }{ \tilde{\psi} (t;t_1,\, t_2 , \dots t_m ; \tilde{W}( \cdot ) ) } \nonumber \\
	\fl
	= \gamma_1^m e^{ - \Gamma  (    t_1  + t_2 + \cdots + t_m  )  }  \bra{\psi_0}   \exp \left[  A (t)^* \ha^\dagger + B^*(t) (\ha^\dagger)^2 \right]  (\ha^\dagger )^m  e^{ -\Gamma t \hn  }   \ha^m \exp \left[  A (t) \ha + B(t) \ha^2 \right]  \ket{\psi_0}  \nonumber \\
	\fl
	= \gamma_1^m e^{ - \Gamma  (    t_1  + t_2 + \cdots + t_m  )  }  \bra{\psi_0} :  e^{ A (t) \ha +  A (t)^* \ha^\dagger + B(t) \ha^2 +  B^*(t) (\ha^\dagger)^2 -    ( 1 -  e^{ - \Gamma t}   ) \ha^\dagger \ha   } (\ha^\dagger \ha)^m  : \ket{\psi_0} ,  \label{totdensity0} 
\end{eqnarray}
where the symbol $:\cdots:$ in equation~(\ref{totdensity0}) indicates normal ordering which places annihilation operators to the right of creation operators.
In deriving the last equality in equation~(\ref{totdensity0}), we used the formula
\begin{equation}
	e^{x \hn} = : e^{\ha^\dagger \ha (e^x -1) } :  \label{normal_exp}
\end{equation}
which is valid for an arbitrary c-number $x$. 
In the limit of $t \rightarrow \infty$, equation~(\ref{totdensity0}) reduces to
\begin{equation}\fl
	  \gamma_1^m e^{ - \Gamma  (    t_1  + t_2 + \cdots + t_m  )  }  \bra{\psi_0} 
	  :  e^{  A (\infty) \ha +  A (\infty)^* \ha^\dagger + B(\infty) \ha^2 +  B^*(\infty) (\ha^\dagger)^2 -    \ha^\dagger \ha   } (\ha^\dagger \ha)^m  :
	   \ket{\psi_0} , \label{totdensity1} 
\end{equation}
where
\begin{eqnarray}
	A(\infty) &=  \sqrt{\gamma_2} \int_0^\infty e^{-\left( i \omega +\frac{\Gamma}{2} \right) t^\prime } \dw (t^\prime) , \\
	B(\infty) &= - \frac{ \gamma_2 }{2i \omega + \Gamma} .
\end{eqnarray}
The joint probability $\tilde{p}_m (t ; \tilde{W} (\cdot ) )$ of $m$-photocounts being recorded during time interval $(0,t)$ 
and the homodyne records $\hrec$ is obtained by integrating (\ref{totdensity0}) 
with respect to $t_1, t_2, \cdots , t_m $ in the integration range $0< t_1 < t_2 < \cdots < t_m <t$.
The relevant part of the integral is the exponential $e^{- \Gamma (t_1 + t_2 + \cdots t_m)}$ 
and evaluated to give
\begin{eqnarray}
	&\int_0^t dt_m   \int_0^{t_m} dt_{m-1} \cdots \int_0^{t_2} dt_1 e^{ - \Gamma  (    t_1  + t_2 + \cdots + t_m  )  } \nonumber \\
	&=  \frac{1}{m!} \left(  \frac{ 1 -  e^{-\Gamma t}   }{\Gamma }  \right)^m .
\end{eqnarray}
Thus, $\tilde{p}_m (t ; \tilde{W} (\cdot ) )$ is given by
\begin{eqnarray} \fl
	\tilde{p}_m (t ; \tilde{W} (\cdot ) ) \nonumber \\ \fl
	=  \bra{\psi_0} : \frac{1}{m!} \left(  \frac{ \gamma_1 }{\Gamma }  (1-e^{-\Gamma t}  ) \ha^\dagger \ha \right)^m  
	e^{  A (t) \ha +  A (t)^* \ha^\dagger + B(t) \ha^2 +  B^*(t) (\ha^\dagger)^2 -    ( 1 -  e^{ - \Gamma t}   ) \ha^\dagger \ha   }
	  :  \ket{\psi_0} \label{normpmw1} \\ \fl
	= \bra{\psi_0}  e^{ A (t)^*  \ha^\dagger + B^*(t) (\ha^\dagger)^2 }   
	\combi{\hn}{m} 
	e^{ -\Gamma t (\hn -m) }  \left(  \frac{ \gamma_1 }{\Gamma }  (1-e^{- \Gamma t}  )  \right)^m   
	e^{     A (t) \ha +B(t) \ha^2 }  \ket{\psi_0},  \label{pdens1}  
\end{eqnarray}
where in deriving the last equality we have used the formula
\begin{eqnarray}
	(\ha^\dagger)^m  e^{x \hn} \ha^m  &=  \hn (\hn -1) \cdots (\hn -m+1)  e^{x(\hn -m)} ,
	\nonumber
\end{eqnarray}
and defined the binomial coefficient operator by
\begin{equation}
	\combi{\hn}{m}
	:=  \frac{\hn (\hn -1) \cdots (\hn -m+1)  }{ m! }.
\end{equation}
In the limit $t \rightarrow \infty$, the joint probability in (\ref{normpmw1}) reduces to
\begin{equation}\fl
	 \bra{\psi_0} : \frac{1}{m!} \left(  \frac{ \gamma_1 }{\Gamma }   \ha^\dagger \ha \right)^m
	\exp \left[  A_\infty \ha +  A_\infty^* \ha^\dagger + B(\infty) \ha^2 +  B^*(\infty) (\ha^\dagger)^2 -   \ha^\dagger \ha   \right]    :  \ket{\psi_0} .
\end{equation}

Note that (\ref{totdensity1}) gives the total probability functional for the measurement outcomes
and that its dependence on the homodyne records enters this formula only through $A (\infty)$.

\subsection{Time Development of Expectation Values of Observables}
For later use, we derive the equation for the conditional time development of expectation value $\mean{\hat{A}}$
of an arbitrary operator $\hA$.

For a one-count process, the density matrix $\hrho (t^+)$ immediately after the state $\hrho (t)$ at time $t$ is given by
$\hrho (t^+) =  \hM_1 \hrho\hM_1^\dagger /\mathrm{tr} [\hM_1 \hrho\hM_1^\dagger]   = \ha \hrho \ha^\dagger / \mean{\hn}$.
Thus, the expectation value $\mean{\hat{A}}_+$ immediately after the photocount event is given by
\begin{equation}
	\mean{\hA}_+ = \mathrm{tr} [ \hrho(t^+) \hA ] = \frac{ \mean{ \ha^\dagger \hA \ha } }{ \mean{\hn} }.   \label{dA_count}
\end{equation}

For a no-count process, the conditional evolution of the density matrix
with homodyne record $\dw$ is given by
\begin{equation}
	\hrho + d \hrho = \frac{ \hM_{\dw} \hrho \hM^\dagger_{\dw}  }{\mean{\hM^\dagger_{\dw} \hM_{\dw}}}.
\end{equation}
Substituting the expression of the measurement operator in (\ref{chmmo1}),
we obtain the equation for the differential of the expectation value $\mean{\hA}$ in our system as follows:
\begin{eqnarray}
	d \mean{\hA}  &=  \frac{ \mathrm{tr} [ \hA  \hM_{ \dw} \hrho  \hM^\dagger_{\dw}  ]  }{  \mathrm{tr} [   \hM_{ \dw } \hrho  \hM^\dagger_{\dw}  ]   } - \mean{\hA} \nonumber \\
	&= \left[ -i \omega \mean{ [ \hA, \hn  ]  } 
	- \frac{\gamma_1}{2} \mean{ \{  \Delta \hA , \Delta \hn   \}  }  
	- \frac{\gamma_2}{ 2 } \mean{  \{ \hA , \hn \}  - 2 \ha^\dagger \hA \ha } \right] dt \nonumber \\
	& \quad + \sqrt{\gamma_2} \mean{ \Delta \hA \Delta \ha  + \Delta \ha^\dagger \Delta \hA  } dW,  \label{dA_nocount}
\end{eqnarray}
where $\Delta \hA := \hA - \mean{\hA}$.

Important examples of the expectation value $\mean{\hA}$
are the average photon number $\mean{\hn}$ and the quadrature amplitude $\mean{\ha}$.
By substituting $\hn$ and $\ha$ into $\hA$, we obtain
for the one-count process
\begin{eqnarray}
	\mean{\hn}_+ &= 
	\frac{ \mean{ \hn^2} - \mean{ \hn  }  }{\mean{\hn}} 
	= \mean{\hn} - 1 + \frac{  \mean{ [\Delta \hn]^2 } }{\mean{\hn}} , 
	\label{dn_count} \\
	\mean{a}_+ &= \frac{  \mean{ \ha^\dagger \ha \ha } }{ \mean{\hn} }
	= \mean{\ha}  + \frac{ \mean{\Delta \hn \Delta  \ha }  }{ \mean{\hn}} ,
	\label{da_count}
\end{eqnarray}
and for the no-count process
\begin{eqnarray}
	d \mean{\hn} &= - \left( \gamma_2 \mean{\hn} + \gamma_1 \mean{[\Delta \hn]^2}     \right) dt
	+ \sqrt{\gamma_2} \mean{ \Delta \hn \Delta \ha + \Delta \ha^\dagger \Delta \hn   } dW, 
	\label{dn_nocount} \\
	d \mean{\ha}  &= - \left[ 
		 \left( i \omega + \frac{\Gamma}{2} \right) \mean{\ha}  
		+ \mean{\Delta \hn \Delta \ha}
	\right] dt
	+ \sqrt{\gamma_2} \mean{  [\Delta \ha]^2 +  \Delta \ha^\dagger \Delta \ha } dW. \label{da_nocount}
\end{eqnarray}

\subsection{Homodyne-record expectation values conditioned on the photon-counting event}
To specify the cross-correlation effects between homodyne and photocount channels,
we derive the expectation values of the homodyne records $\dw (t \mp 0)$
immediately before/after a photodetection at time $t$.
The expectation value of the homodyne record $\dw (t)$
for the system's wave function $\ket{ \psi(t)}$
is given by
\begin{equation}
	E[ \dw (t)|\rho(t) ] = \sqrt{\gamma_2} \mean{ \ha + \ha^\dagger) } dt.
\end{equation}
Thus, we have only to compare the quantum expectations of the quadrature amplitude
$\mean{\ha + \ha^\dagger}_{\mp}$ immediately before/after the photon-counting.
From Eq.~(\ref{conpsi3}) the unnormalized conditional wave function at time $t$ is, up to a multiplicative c-number factor, given by
\begin{equation}
	\ket{\bar{\psi}_m(t;\tilde{A}(t), \tilde{B}(t))} = \ha^m \exp \left[  \tilde{A} (t) \ha +   \tilde{B} (t) \ha^2 \right] e^{ - \left(i \omega + \frac{\Gamma}{2}\right) t \hn } \ket{\psi_0}.
	\label{conpsi4}
\end{equation}
Thus, we obtain
\begin{eqnarray}
	\fl
	\mean{\ha + \ha^\dagger}_- 
	\nonumber \\ \fl
	= \frac{ \bra{\bar{\psi}_m(t;\tilde{A}^* (t), \tilde{B}^*(t))} (\ha + \ha^\dagger) \ket{\bar{\psi}_m(t;\tilde{A}(t), \tilde{B}(t))} }{
	\braket{\bar{\psi}_m(t;\tilde{A}^*(t), \tilde{B}^*(t))}{\bar{\psi}_m(t;\tilde{A}(t), \tilde{B}(t))} } \nonumber
	\\ \fl
	= (\partial_{\tilde{A} }  + \partial_{\tilde{A}^*}) \log \left.
	[ \braket{\bar{\psi}_m(t;\tilde{A}^*, \tilde{B}^*)}{ \bar{\psi}_m(t;\tilde{A}, \tilde{B})  }]
	\right|_{ \tilde{A} = \tilde{A}(t), \, \tilde{B} =\tilde{B}(t) } 
	\nonumber \\ \fl
	= (\partial_{\tilde{A} }  + \partial_{\tilde{A}^*}) \log \left.
	[ \partial_{\tilde{A}}^m \partial_{\tilde{A}^*}^m
	 \braket{\bar{\psi}_0(t;\tilde{A}^*, \tilde{B}^*)}{ \bar{\psi}_0(t;\tilde{A}, \tilde{B})  }]
	\right|_{ \tilde{A} = \tilde{A}(t), \, \tilde{B} =\tilde{B}(t) } ,
	\label{condexp2} \\ \fl
	\mean{\ha + \ha^\dagger}_+
	\nonumber \\ \fl
	= \frac{ \bra{\bar{\psi}_m(t;\tilde{A}^* (t), \tilde{B}^*(t))} 
	\ha^\dagger (\ha + \ha^\dagger) \ha
	\ket{\bar{\psi}_m(t;\tilde{A}(t), \tilde{B}(t))} }{
	\bra{\bar{\psi}_m(t;\tilde{A}^*(t), \tilde{B}^*(t))} \ha^\dagger \ha \ket{\bar{\psi}_m(t;\tilde{A}(t), \tilde{B}(t))} }
	\nonumber \\ 
	\fl
	= (\partial_{\tilde{A} }  + \partial_{\tilde{A}^*}) \log \left.
	[ \partial_{\tilde{A}} \partial_{\tilde{A}^*}
	\braket{\bar{\psi}_m(t;\tilde{A}^*, \tilde{B}^*)}{ \bar{\psi}_m(t;\tilde{A}, \tilde{B})  }]
	\right|_{ \tilde{A} = \tilde{A}(t), \, \tilde{B} =\tilde{B}(t) } 
	\nonumber\\
	\fl
	= (\partial_{\tilde{A} }  + \partial_{\tilde{A}^*}) \log \left.
	[ \partial_{\tilde{A}}^{m+1} \partial_{\tilde{A}^*}^{m+1}
	 \braket{\bar{\psi}_0(t;\tilde{A}^*, \tilde{B}^*)}{ \bar{\psi}_0(t;\tilde{A}, \tilde{B})  }]
	\right|_{ \tilde{A} = \tilde{A}(t), \, \tilde{B} =\tilde{B}(t) } ,
	\label{condexp4} 
\end{eqnarray}
where $\partial_x := \partial / \partial x$.
Note that the bra vector $\bra{\bar{\psi}_m(t;\tilde{A}^* (t), \tilde{B}^*(t))}$ conjugate
to the ket vector in (\ref{conpsi4})
depends on $\tilde{A}^*$ and $\tilde{B}^*$.
Equations~(\ref{condexp2}) and (\ref{condexp4}) quantitatively express
the back action of the photon-counting measurement to the homodyne records.
We note that the expression (\ref{condexp4}) involves additional derivatives with respect to $\tilde{A}$ and $\tilde{A}^*$
which reflect the effect of one photocount as can be
seen from (\ref{conpsi4}) (see also (\ref{conpsi3})).
\subsection{Generating functional}
In this subsection, we derive a general formula for the generating functional of measurement records $\dw_t$ and $dN_t$,

Instead of deriving the generating functional of $\dw_t$ and $dN_t$,
we discuss that of $dA (t) = \sqrt{\gamma_2}  e^{- \left( \frac{\Gamma}{2} + i \omega  \right) t } \dw_t$ and $dN_t$,
which is defined as
\begin{equation}
	M[\xi (\cdot) , \xi^* (\cdot)  ,  \eta (\cdot) ] = E \left[ 
	e^{  \int_0^\infty \xi (t^\prime)  dA (t^\prime)  +   \int_0^\infty \xi^* (t^\prime)  dA^*_{t^\prime}   +  \int_0^\infty \eta (t^\prime) dN_{t^\prime}     }
	\right] ,   \label{defgenfunc1}
\end{equation}
where $\xi$ and $\eta$ are arbitrary functions of $t$.
This functional contains all the information about the probability distribution of measurement records.

To calculate this generating functional,
we note that
the stochastic integral $ \int_0^\infty \eta (t^\prime) dN_{t^\prime}$ becomes $\sum_{k=1}^m  \eta (t_k)$,
if the photocounts occur at times $t_1< t_2 < \cdots < t_m$.
Thus, the generating functional can be evaluated as
\begin{eqnarray} \fl
	M[\xi (\cdot) ,\xi^* (\cdot) ,\eta (\cdot) ] 
	=  \sum_{m=0}^\infty \int_0^\infty  dt_m \int_0^{t_m} dt_{m-1} \cdots \int_0^{t_2} dt_1  
	\int \mu_0 (\tilde{W} (\cdot )) \nonumber \\
	\fl
	\times \braket{ \tilde{\psi}_\infty (t_1 , t_2 , \cdots , t_m  ; \tilde{W} (\cdot ) )   }{\tilde{\psi}_\infty  (t_1 , t_2 , \cdots , t_m  ; \tilde{W} (\cdot) )}  
	\nonumber \\
	\fl \times
	\exp \left[  \int_0^\infty \xi (t^\prime)  dA (t^\prime) +   \int_0^\infty \xi^* (t^\prime)  dA^*(t^\prime)  +  \sum_{k=0}^m  \eta (t_k)   \right] .
\end{eqnarray}
From the square norm of the wave function in Eq. (\ref{totdensity1}), we obtain
\begin{eqnarray}
	\fl M[\xi (\cdot) , \xi^* (\cdot ), \eta (\cdot) ] \nonumber \\
	\fl = \sum_{m=0}^\infty \int_0^\infty  dt_m \int_0^{t_m} dt_{m-1} \cdots \int_0^{t_2} dt_1    \int \mu_0 (\tilde{W} (\cdot ))  \gamma_1^m e^{ - \Gamma (t_1 + t_2 + \cdots + t_m) +  \sum_{k=0}^m  \eta (t_k)  } \nonumber \\
	\fl \times \bra{ \psi_0  }   :  e^{    \int_0^\infty \xi (t^\prime)  dA (t^\prime) +   \int_0^\infty \xi^* (t^\prime)  dA^* (t^\prime)  
	+   A_\infty  \ha + A_\infty^* \ha^\dagger 
	+  B(\infty)  \ha^2  + B^* (\infty)  (\ha^\dagger )^2      -   \ha^\dagger \ha   }  (  \ha^\dagger \ha  )^m      :  \ket{ \psi_0  }   \nonumber \\
	\fl =  \bra{\psi_0}  :  \sum_{m=0}^\infty \frac{ 1 }{m!} \left( \gamma_1 \ha^\dagger \ha  \int_0^\infty e^{ - \Gamma t^\prime + \eta (t^\prime) } dt^\prime  \right)^m \int \mu_0  (\tilde{W} (\cdot) )  \nonumber \\
	\fl \times  
	e^{  \sqrt{\gamma_2} \int_0^\infty \dw_{t^\prime}  [ e^{- \left( i\omega +  \frac{\Gamma }{2}  \right) t^\prime }  ( \xi (t^\prime )  + \ha ) +  e^{- \left( - i \omega +  \frac{\Gamma }{2}  \right) t^\prime }  ( \xi^* (t^\prime )  + \ha^\dagger ) ]
	+  B(\infty)  \ha^2  + B^* (\infty)  (\ha^\dagger )^2  - \ha^\dagger \ha   }  :  \ket{\psi_0}  \nonumber \\
    \fl = \exp \left[  \frac{\gamma_2}{2}  \int_0^\infty | \xi (t^\prime) e^{- \left(i\omega + \frac{ \Gamma}{2} \right) t^\prime  } + \xi^* (t^\prime) e^{- \left(- i\omega + \frac{ \Gamma}{2} \right) t^\prime  }   |^2 dt^\prime \right]
  \bra{\psi_0} : e^{ \kappa \ha + \kappa^* \ha^\dagger + \nu \ha^\dagger \ha   }   :  \ket{\psi_0} ,
 \label{charfun1}
\end{eqnarray}
where
\begin{eqnarray}
	\kappa &= \gamma_2 \int_0^\infty  ( e^{- (2i\omega +  \Gamma) t^\prime  }    \xi (t^\prime) 
	+ e^{- \Gamma  t^\prime  }  \xi^* (t^\prime)  )  dt^\prime , \label{defkappa} \\
	\nu &= \gamma_1 \int_0^\infty e^{- \Gamma t^\prime }  ( e^{\eta (t^\prime)}  -1  ) dt^\prime .  \label{defnu}
\end{eqnarray}
Equation (\ref{charfun1}) gives the general formula for the generating functional of measurement records.

The generating function $E[e^{\xi A (t) + \xi^* A (t)^* + \eta N_t}]$ with respect to output variables $A (t)$ and $N_t$
can be derived by substituting
\begin{eqnarray}
	\xi (t^\prime)  &=  \left\{ \begin{array}{l} \xi  \quad (0 < t^\prime < t); \\ 0 \quad (\mathrm{otherwise}),   \end{array} \right.
	\\
	\eta (t^\prime)  &= \left\{ \begin{array}{l} \eta \quad (0 < t^\prime < t);  \\ 0 \quad (\mathrm{otherwise})   \end{array}\right.
\end{eqnarray}
into Eq.~(\ref{charfun1}).
The result takes a form similar to Eq.~(\ref{charfun1}):
\begin{eqnarray}
	&E[e^{\xi A (t) + \xi^* A (t)^* + \eta N_t}] \nonumber \\
	&=  \exp \left[  \frac{\gamma_2}{2}   
	\left( \frac{1- e^{- (2i\omega + \Gamma) t^\prime }}{ 2i\omega + \Gamma  } \xi^2 
	 + \frac{ 1- e^{-(- 2i\omega + \Gamma) t }}{-2i\omega + \Gamma} (\xi^*)^2  +  2 \frac{1- e^{-\Gamma t}}{\Gamma} |\xi|^2  \right) \right] 
	\nonumber  \\ & \quad \times 
	\bra{\psi_0} : \exp[ \kappa_t \ha + \kappa_t^* \ha^\dagger + \nu_t \ha^\dagger \ha   ]   :  \ket{\psi_0} , \label{charfun2} 
\end{eqnarray}
where
\begin{eqnarray}
	\kappa_t &= \gamma_2 \left(  \frac{1- e^{-( 2i\omega + \Gamma) t^\prime }}{ 2i\omega + \Gamma  } \xi + \frac{1- e^{-\Gamma t}}{\Gamma} \xi^*   \right) ,
	\label{defkappat} \\
	\nu_t &= \frac{\gamma_1}{\Gamma}  (1-e^{-\Gamma t}) (e^\eta -1).  \label{defnut}
\end{eqnarray}

To gain the physical insights of the formulas (\ref{charfun1}) and (\ref{charfun2}), let us assume that $\ha$ and $\ha^\dagger$ are c-numbers.
Then, the generating functional (\ref{charfun1}) would be that of independent stochastic processes, $\dw (t)$ and $dN (t)$.
Such a description is justified only when the initial state is a coherent state (see Sec.~4); 
otherwise, there will, in general, 
be correlations between these output records 
because of the noncommutativity of $\ha$ and $\ha^\dagger$.
In this sense, 
the measurement records give us information 
about the system's deviation from the coherent state
as exemplified in the next section.
From a standpoint of measurement theory,
the correlation between the two output records,
which can be seen in the generating functional~(\ref{charfun1}),
reflect the backaction of one measurement channel on the other,
thus revealing the wave-particle duality in the measurement process, that is, the measurement backaction of photodetection influences the output records of homodyne detection, and vice versa.

\section{Application to typical initial conditions}
In this section we apply the general formulas obtained in the previous sections to typical quantum states:
coherent, number, thermal, and squeezed states.

\subsection{Coherent state}
The coherent state $\ket{\alpha}$ is represented in the number-state basis as
\[
	\ket{\alpha } =   e^{- \frac{ | \alpha |^2  }{2}} \sum_{n=0}^\infty \frac{\alpha^n}{\sqrt{n!}} \ket{n}    
\]
where $\alpha$ is an arbitrary complex number.
The coherent state is an eigenstate of the boson annihilation operator:
\begin{equation}
	\ha \ket{\alpha} = \alpha \ket{\alpha}.  \label{coherenteigenrelation}
\end{equation}

The conditional wave function of the $m$-count process in (\ref{conpsi3}) for an initial state $\ket{\psi_0} = \ket{\alpha} $ is given by
\begin{eqnarray}
	\fl
	\ket{ \tilde{\psi} (t;t_1,\, t_2 , \dots t_m ; \tilde{W}( \cdot ) )  }  \nonumber \\
	\fl = \gamma_1^{m/2}\alpha^m
	\exp\left[  - \left(i \omega + \frac{\Gamma}{2}\right)  (    t_1  + t_2 + \cdots + t_m  ) 
	+ A (t) \alpha +   B(t) \alpha^2 - \frac{|\alpha |^2}{2}(1 - e^{- \Gamma t}) \right]
	\nonumber \\ \fl
	\times
	 \ket{ e^{ - \left(i \omega + \frac{\Gamma}{2}\right) t \hn } \alpha } . \label{mcountcoherent} 
\end{eqnarray}
In evaluating the $m$-count wave function, 
we used (\ref{coherenteigenrelation}) and the formula
\begin{equation}
	e^{\lambda \hn}  \ket{\alpha} =  e^{-  \frac{|\alpha |^2}{2} ( 1- |e^\lambda|^2 )  }  \ket{e^\lambda \alpha} .  \label{e^lcoherent}
\end{equation}
It follows from Eq.~(\ref{mcountcoherent}) 
that the normalized state vector does not depend
on the measurement outcomes and is given by 
$ \ket{ e^{ - \left(i \omega + \frac{\Gamma}{2}\right) t \hn } \alpha }$; 
the system develops deterministically.

The generating functional in (\ref{charfun1}) is evaluated as 
\begin{eqnarray} \fl
	M[\xi(\cdot), \xi^*(\cdot), \eta (\cdot)] 
	\nonumber \\ \fl
	= \exp \left[  \frac{\gamma_2}{2}  \int_0^\infty | \xi (t^\prime) e^{- \left(i\omega + \frac{ \Gamma}{2} \right) t^\prime  } + \xi^* (t^\prime) e^{- \left(- i\omega + \frac{ \Gamma}{2} \right) t^\prime  }   |^2
	dt^\prime 
	+\kappa \alpha + \kappa^* \alpha^* + \nu |\alpha |^2
	\right] ,   \label{charfuncoherent}
\end{eqnarray}
where we used the formula
\begin{equation}
	\bra{\alpha} : f(\ha, \ha^\dagger) : \ket{\alpha}  
	= f (\alpha, \alpha^*) . \label{normalexcoherent}
\end{equation}
Note that the generating functional in (\ref{charfuncoherent}) can be obtained by replacing $\ha$ and $\ha^\dagger$ 
by the corresponding c-numbers, $\alpha $ and $\alpha^*$.
From the definitions of $\kappa$ and $\nu$ in equations (\ref{defkappa}) and (\ref{defnu}),
we find from (\ref{charfuncoherent}) that
\begin{equation}
	M[\xi(\cdot), \xi^* (\cdot), \eta (\cdot)] = M[\xi(\cdot), \xi^* (\cdot),  0]
	 \times M[  0, 0,  \eta (\cdot )],
\end{equation}
which implies that homodyne records $\hrec$ and photocount records are statistically independent.

From Eq.~(\ref{charfun2}), the generating function of $A (t)$ is
\begin{eqnarray}
	\fl E[e^{\xi A (t) + \xi^* A (t)^*}] \nonumber \\
	\fl  =  \exp \left[  \frac{\gamma_2}{2}   
	\left( \frac{1- e^{ (2i\omega + \Gamma) t^\prime }}{ 2i\omega + \Gamma  } \xi^2 
	 + \frac{ 1- e^{-(- 2i\omega + \Gamma) t }}{-2i\omega + \Gamma} (\xi^*)^2  +  2 \frac{1- e^{-\Gamma t}}{\Gamma} |\xi|^2  \right) 
	\right.  \nonumber \\ \fl  \quad \quad  \left.
	+ \xi \gamma_2 \left(  \alpha \frac{1- e^{ (2i\omega + \Gamma) t^\prime }}{ 2i\omega + \Gamma  } + \alpha^* \frac{ 1 - e^{- \Gamma t} }{\Gamma}   \right)   
	\right.  \nonumber \\ \fl  \quad \quad  \left.
	+ \xi^* \gamma_2 \left(  \alpha^* \frac{1- e^{ (-2i\omega + \Gamma) t^\prime }}{- 2i\omega + \Gamma  } + \alpha \frac{ 1 - e^{- \Gamma t} }{\Gamma}   \right)  \right] ,
\end{eqnarray}
which implies that $A (t)$ is a complex Gaussian variable with its first and second moments given as follows:
\begin{eqnarray}
	&E[A (t)] = \gamma_2 \left(  \alpha \frac{1- e^{ (2i\omega + \Gamma) t^\prime }}{ 2i\omega + \Gamma  } + \alpha^* \frac{ 1 - e^{- \Gamma t} }{\Gamma} \right) , \\
	&E[(A (t) - E[A (t)] )^2  ] = \frac{\gamma_2}{2} \frac{1- e^{ (2i\omega + \Gamma) t^\prime }}{ 2i\omega + \Gamma  }, \\
	&E[| A (t) - E[A (t)] |^2]  =  \gamma_2 \frac{1- e^{-\Gamma t}}{\Gamma}.
\end{eqnarray}

\subsection{Number state}
\begin{figure}
\centering
\includegraphics[width=10cm,clip]{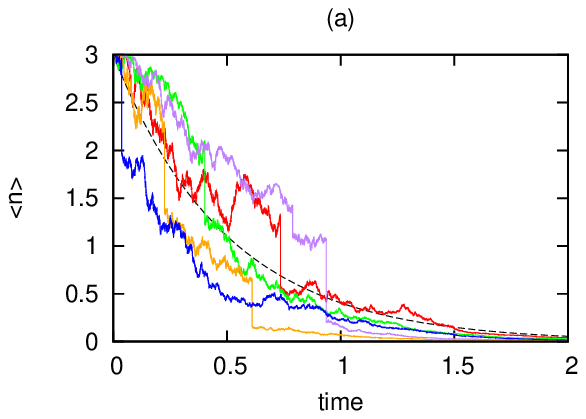}
\includegraphics[width=10cm,clip]{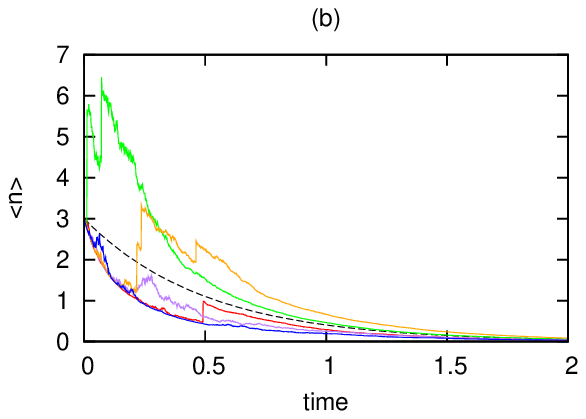}
\includegraphics[width=10cm,clip]{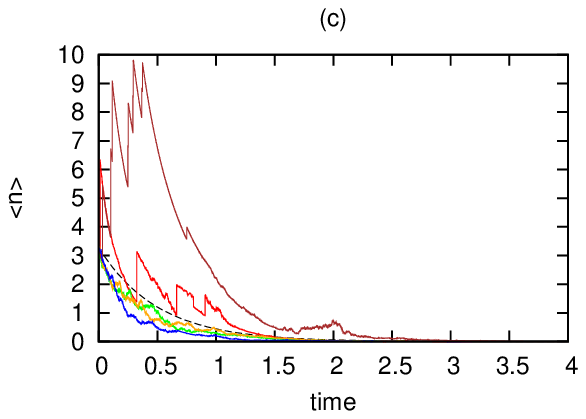}
\caption{(Color online)
	Monte Carlo paths of $\mean{\hn}$ starting from
	(a) number, (b) thermal and (c) squeezed states.
	In each figure, the dashed curve shows the time evolution of
	the ensemble average of the photon number which is given by $\mean{\hn}_0 e^{- \Gamma t}$,
	where the subscript 0 indicates the average over the initial state.
	The parameters used are $\gamma_1 = \gamma_2 = 1$ and $\omega = 0$.
	For the number and thermal paths, 
	the initial photon number is  $\mean{\hn}_0 = 3$.
	The parameters for the initial squeezed state are $r = 1.2,$ and $\alpha = 1.$
	For the case of (a),
	the change in $\mean{\hn}$ upon photodetection
	is negative,
	while for (b) and (c), it is positive.
	This reflects the sub-Poissonian photon number distributions in (a) and
	the super-Poissonian distributions in (b) and (c).
	There also appears diffusive behavior in the no-count processes 
	arising from homodyne detection
	in all of the three cases.
}
\label{fig_num1}
\end{figure}
The $m$-count wave function for an initial number state $\ket{\psi_0} = \ket{n}$ is evaluated as
\begin{eqnarray}
	\fl  \ket{ \tilde{\psi} (t;t_1,\, t_2 , \dots t_m ; \tilde{W}( \cdot ) )  } \nonumber \\
	\fl =\gamma_1^{m/2} e^{   - \left(i \omega + \frac{\Gamma}{2}\right)  (    t_1  + t_2 + \cdots + t_m  ) } \nonumber \\
	\fl \quad \times \sum_{k=0}^{n-m} \sum_{l=0}^{\lfloor k/2 \rfloor}  \frac{ A (t)^{k-2l} B(t)^l }{(k-2l)! l!} \sqrt{\frac{n!}{(n-m-k)!}} 
	e^{- \left( i \omega + \frac{\Gamma}{2}  \right) (n-m-k) t }\ket{n-m-k} ,
\end{eqnarray}
where $\lfloor x \rfloor $ is the largest integer that does not exceed $x$.
Note that $m \leq n$ since $\ket{ \tilde{\psi} (t;t_1,\, t_2 , \dots t_m ; \tilde{W}( \cdot ) )  }$
vanishes for $m>n$.

The joint distribution function $\tilde{p}_m (t ; \tilde{W} (\cdot ) )$ in (\ref{totdensity0}) becomes
\begin{eqnarray}
	\fl \tilde{p}_m (t ; \tilde{W} (\cdot ) ) \nonumber \\
	\fl =  \left(  \frac{ \gamma_1 }{\Gamma }  (1-e^{- \Gamma t}  )  \right)^m
	||  e^{- (i\omega + \frac{\Gamma}{2}) t \hn  }  \exp[A (t) \ha +  B(t) \ha^2  ]  \ha^m  \ket{n}  ||^2   \nonumber \\
	\fl = \left(  \frac{ \gamma_1 }{\Gamma }  (1-e^{- \Gamma t}  )  \right)^m
	\sum_{k=0}^{n-m} \left| \sum_{l=0}^{\lfloor k/2 \rfloor}   \frac{ A (t)^{k-2l} B(t)^l }{(k-2l)! l!}    \right|^2 \frac{n!}{(n-m-k)!} 
	e^{- \Gamma t (n-m-k)  } .  \label{densitynumber1}
\end{eqnarray}
In the limit of $t \rightarrow \infty$, (\ref{densitynumber1}) reduces to
\begin{eqnarray}
	\tilde{p}_m (\infty ; \tilde{W} (\cdot ) ) =  n! \left(  \frac{ \gamma_1 }{\Gamma }   \right)^m
	    \left| \sum_{l=0}^{\lfloor (n-m)/2 \rfloor}   \frac{ A(\infty)^{k-2l}  B(\infty)^l }{(n-m-2l)! l!}    \right|^2 .
\end{eqnarray}

The generating functional takes a simpler form.
The calculation proceeds as follows:
from
\begin{eqnarray}
	e^{\kappa \ha} \ket{n} =  \sum_{m=0}^n  \frac{\kappa^m}{m!} \sqrt{\frac{n!}{(n-m!)}} \ket{n-m} ,
\end{eqnarray}
we have
\begin{eqnarray}\fl
	\bra{n}: \exp [ \kappa \ha +  \kappa^* \ha^\dagger +  \nu \ha^\dagger \ha ]  : \ket{n} 
	&=  ( e^{\kappa \ha} \ket{n})^\dagger  (1+\nu )^{\hn}  e^{\kappa \ha } \ket{n}  \nonumber \\
	&= \sum_{m=0}^n  \frac{n!}{(m!)^2  (n-m)!} |\kappa |^{2m} (1+\nu )^{n-m} \nonumber \\
	&=  (1+ \nu )^n L_n \left(- \frac{ | \kappa |^2 }{1+ \nu} \right) ,
\end{eqnarray}
where $L_n (x) $ is the Laguerre polynomial defined by
\begin{eqnarray}
	L_n (x) := \frac{e^x}{n!} \frac{d^n}{dx^n} (e^{-x} x^n) = \sum_{m=0}^n \frac{n!}{(m!)^2  (n-m)!} (-x)^m . \label{deflaguerre}
\end{eqnarray}
Thus, the generating functional and the generating function are given by
\begin{eqnarray}
	\fl M[\xi(\cdot),  \xi^* (\cdot), \eta (\cdot )]  \nonumber \\
	\fl= \exp \left[  \frac{\gamma_2}{2}  \int_0^\infty | \xi (t^\prime) e^{- \left(i\omega 
	+ \frac{ \Gamma}{2} \right) t^\prime  } + \xi^* (t^\prime) e^{- \left(- i\omega + \frac{ \Gamma}{2} \right) t^\prime  }   |^2 dt^\prime \right]
	(1+ \nu )^n L_n \left(- \frac{ | \kappa |^2 }{1+ \nu} \right) , \\
	\fl E[e^{\xi A (t) + \xi^* A (t)^* + \eta N_t}] \nonumber \\
	\fl =  \exp \left[  \frac{\gamma_2}{2}   
	\left( \frac{1- e^{ -(2i\omega + \Gamma) t^\prime }}{ 2i\omega + \Gamma  } \xi^2 
	 + \frac{ 1- e^{-(- 2i\omega + \Gamma) t }}{-2i\omega + \Gamma} (\xi^*)^2  +  2 \frac{1- e^{-\Gamma t}}{\Gamma} |\xi|^2  \right) \right] 
	\nonumber  \\ \fl \quad \times 
	(1+ \nu_t )^n L_n \left(- \frac{ | \kappa_t |^2 }{1+ \nu_t} \right) .
\end{eqnarray}
Note that for the initial number state the generating function(al) cannot be factorized
into a function of $\xi$ and that of $\eta$, 
implying that there are correlations between these two measurement outcomes.

The Monte Carlo simulation for the measurement process is done 
for the number-state initial condition.
The simulation method is as follows:
For each time step $\Delta t$,
we first check if the photodetection occurs or not.
If it does, 
the state vector or the density operator evolves according to the jump operator (\ref{chmmo2}).
If not, the state evolves according to the diffusive measurement operator (\ref{chmmo1}) 
corresponding to homodyne detection.
 
The results for an initial number state $\ket{n}$ with $n=3$ 
are shown in figure~\ref{fig_num1} (a).
In the no-count event,
the expectation value of the photon number 
decreases on average according to (\ref{dn_nocount}),
while there are local stochastic deviations which arise from the diffusive term
$ \sqrt{\gamma_2} \mean{ \Delta \hn \Delta \ha + \Delta \ha^\dagger \Delta \hn   } dW$
in (\ref{dn_nocount}).

\subsection{Thermal state}

Let us assume now that the initial state is a thermal state
\begin{eqnarray}
	\hrho_0 = \frac{ e^{- \beta \omega \hn}  }{ Z  } 
	= (1-e^{-\beta \omega}) \sum_{n=0}^\infty e^{-\beta \omega n} \ket{n} \bra{n} , \nonumber 
\end{eqnarray}
where
\begin{eqnarray}
	Z = \mathrm{tr} [ e^{-\beta \omega \hn} ] = \frac{1}{ 1- e^{-\beta \omega} }, \nonumber
\end{eqnarray}
and $\beta$ is the inverse temperature.
Then, $\tilde{p}_m (t; \tilde{W} (\cdot ) )$ can be calculated
by taking the ensemble average 
of the corresponding quantity 
for the initially number state over $n$:
\begin{eqnarray}
	\fl  \tilde{p}_m (t ; \tilde{W} (\cdot ) )  \nonumber \\ 
	\fl  = \frac{ 1 }{Z} \sum_{n=m}^\infty e^{-\beta \omega n}  \left(  \frac{ \gamma_1 }{\Gamma }  (1-e^{- \Gamma t}  )  \right)^m
	\sum_{k=0}^{n-m} \left| \sum_{l=0}^{\lfloor k/2 \rfloor}   \frac{ A (t)^{k-2l} B(t)^l }{(k-2l)! l!}    \right|^2 \frac{n!}{(n-m-k)!} 
	e^{- \Gamma t (n-m-k)  } \nonumber \\
	\fl = \frac{ e^{- \beta \omega m} }{Z}  \left(  \frac{ \gamma_1 }{\Gamma }  (1-e^{- \Gamma t}  )  \right)^m
	 \sum_{k=0}^{\infty} \left| \sum_{l=0}^{\lfloor k/2 \rfloor}   \frac{ A (t)^{k-2l} B(t)^l }{(k-2l)! l!}    \right|^2
	\frac{ e^{- \beta \omega k} (k+m)! }{( 1 - e^{-(\Gamma t + \beta \omega )})^{k+m+1}},
\end{eqnarray}
where in the last equality the following formula was used:
\begin{eqnarray}
	\sum_{n=0}^\infty \frac{(n+m)!}{(n-k)!} x^n = \frac{x^k (k+m)!}{(1-x)^{m+k+1}}. \label{kousiki1}
\end{eqnarray}

The generating functional and the generating function are evaluated as follows:
\begin{eqnarray}
	\fl  M[\xi(\cdot),  \xi^* (\cdot), \eta (\cdot )]  \nonumber \\
	\fl =   \exp \left[  \frac{\gamma_2}{2}  \int_0^\infty | \xi (t^\prime) e^{- \left(i\omega 
	+ \frac{ \Gamma}{2} \right) t^\prime  } + \xi^* (t^\prime) e^{- \left(- i\omega + \frac{ \Gamma}{2} \right) t^\prime  }   |^2 dt^\prime \right]
	\sum_{n=0}^\infty \frac{e^{-\beta \omega n} }{Z} (1+ \nu )^n L_n \left(- \frac{ | \kappa |^2 }{1+ \nu} \right) \nonumber \\
	\fl = \frac{e^{\beta \omega} -1 }{e^{\beta \omega} - 1 - \nu } 
	\exp \left[ \frac{|\kappa |^2}{e^{\beta \omega} -1-\nu} + \frac{\gamma_2}{2}  \int_0^\infty | \xi (t^\prime) e^{- \left(i\omega 
	+ \frac{ \Gamma}{2} \right) t^\prime  } + \xi^* (t^\prime) e^{- \left(- i\omega + \frac{ \Gamma}{2} \right) t^\prime  }   |^2 dt^\prime \right] ,
	\label{charfunth1} \\
	\fl E[e^{\xi A (t) + \xi^* A (t)^* + \eta N_t}] \nonumber \\
	\fl = \frac{e^{\beta \omega} -1 }{e^{\beta \omega } - 1 - \nu_t }  \exp \left[ \frac{|\kappa_t |^2}{e^{\beta \omega } -1-\nu_t} \right. \nonumber \\
	\fl  \left. \quad+ \frac{\gamma_2}{2}   \left( \frac{1- e^{- (2i\omega + \Gamma) t^\prime }}{ 2i\omega + \Gamma  } \xi^2 
	+ \frac{ 1- e^{-(- 2i\omega + \Gamma) t }}{-2i\omega + \Gamma} (\xi^*)^2  +  2 \frac{1- e^{-\Gamma t}}{\Gamma} |\xi|^2  \right) \right] .
	\label{charfunth2}
\end{eqnarray}
In deriving the last equality in (\ref{charfunth1}), we used the relation
\begin{equation}
	\sum_{n=0}^\infty t^n L_n (x) =\frac{1}{1-t} \exp \left[ - \frac{xt}{1-t}  \right].
\end{equation}
Again, these characteristic functions are not separable
with respect to $\xi$ and $\eta$,
reflecting the mutual influence between two measurement outcomes.

The Monte Carlo paths of $\mean{\hn}$ is shown in figure~\ref{fig_num1} (b).
The average behavior showing an exponential damping in time 
is the same as the number state,
while the change in the average photon number upon photodetection is positive,
reflecting the fact that the photon number distribution is super-Poissonian
(see equation~(\ref{dn_nocount}))~\cite{ueda_ogawa}.

\subsection{Squeezed state}

Finally, we consider the case 
in which a squeezed state is taken as the initial condition:
\begin{eqnarray}
	\ket{\alpha , r} & := D(\alpha ) S(r) \ket{0}  , \label{squeezed} 
\end{eqnarray}
where
\begin{eqnarray}
	D( \alpha  )  &:= e^{\alpha \ha^\dagger  -  \alpha^*  \ha},  \\
	S(r) &:= e^{\frac{r}{2} (\ha^2 - \ha^{\dagger  2}) } ,
\end{eqnarray}
with $\alpha $ being an arbitrary complex number and $r$ an arbitrary real number which is called a squeezing parameter.

To evaluate the generating function(al), we need to calculate
\begin{eqnarray}
	\bra{\alpha , r} :  \exp [  \kappa \ha +  \kappa^* \ha^\dagger + \nu \ha^\dagger \ha ]  : \ket{\alpha , r} .  \label{sqchar1}
\end{eqnarray}
This is done in Appendix A with the result
\begin{eqnarray}
	&\left[  
	\left( 1+ \frac{\nu}{2}(1-e^{-2r})  \right) \left( 1+ \frac{\nu}{2}(1-e^{2r})  \right)  
	\right]^{-1/2}
	\nonumber \\
	&\times \exp \left[     - \frac{ | \kappa |^2  }{1+\nu}  +  \frac{1}{ \frac{2}{1+ e^{-2r}}  -  \frac{\nu}{1+\nu}  }  \left( \frac{\mathrm{Re} \kappa}{1+ \nu}  + \frac{2 \mathrm{Re} \alpha}{ 1+ e^{-2r} } \right)^2 - \frac{ 2 (\mathrm{Re} \alpha )^2 }{1+ e^{-2r}}  
	 \right.  \nonumber \\ & \left. 
	+ \frac{1}{ \frac{2}{1+ e^{2r}}  -  \frac{\nu}{1+\nu}  }  \left( \frac{ - \mathrm{Im} \kappa}{1+ \nu}  + \frac{2 \mathrm{Im} \alpha}{ 1+ e^{2r} } \right)^2  -    \frac{ 2 (\mathrm{Im} \alpha )^2}{1+ e^{2r}}    \right] .  \label{sqchar2}
\end{eqnarray}
Thus, the generating functional and the generating function are given by
\begin{eqnarray}
	 \fl M[\xi(\cdot),  \xi^* (\cdot), \eta (\cdot )]  \nonumber \\
	\fl =  \left[ 
	\left( 1+ \frac{\nu}{2}(1-e^{-2r})  \right) \left( 1+ \frac{\nu}{2}(1-e^{2r})  \right)  
	\right]^{-1/2}
	\nonumber \\
	\fl \quad \times \exp \left[  \frac{\gamma_2}{2}  \int_0^\infty | \xi (t^\prime) e^{- \left(i\omega 
	+ \frac{ \Gamma}{2} \right) t^\prime  } + \xi^* (t^\prime) e^{- \left(- i\omega + \frac{ \Gamma}{2} \right) t^\prime  }   |^2 dt^\prime 
	 \right.  \nonumber \\ \fl   
	- \frac{ | \kappa |^2  }{1+\nu}  +  \frac{1}{ \frac{2}{1+ e^{-2r}}  -  \frac{\nu}{1+\nu}  }  \left( \frac{\mathrm{Re} \kappa}{1+ \nu}  + \frac{2 \mathrm{Re} \alpha}{ 1+ e^{-2r} } \right)^2 
	+ \frac{1}{ \frac{2}{1+ e^{2r}}  -  \frac{\nu}{1+\nu}  }  \left( \frac{ - \mathrm{Im} \kappa}{1+ \nu}  + \frac{2 \mathrm{Im} \alpha}{ 1+ e^{2r} } \right)^2 
	\nonumber \\  
	\fl \left. 
	- \frac{ 2 (\mathrm{Re} \alpha )^2 }{1+ e^{-2r}} 
	 -    \frac{ 2 (\mathrm{Im} \alpha )^2}{1+ e^{2r}}  \right]	,   \label{sqchrfunc1} \\
	\fl E[e^{\xi A (t) + \xi^* A (t)^* + \eta N_t}] \nonumber \\
	\fl =
	\left[  
	\left( 1+ \frac{\nu}{2}(1-e^{-2r})  \right) \left( 1+ \frac{\nu}{2}(1-e^{2r})  \right)  
	\right]^{-1/2}
	\nonumber \\
	\fl  \quad
	\times \exp \left[  \frac{\gamma_2}{2}  \left( \frac{1- e^{- (2i\omega + \Gamma) t^\prime }}{ 2i\omega + \Gamma  } \xi^2 
	 + \frac{ 1- e^{-(- 2i\omega + \Gamma) t }}{-2i\omega + \Gamma} (\xi^*)^2  +  2 \frac{1- e^{-\Gamma t}}{\Gamma} |\xi|^2  \right)
	  \right.  \nonumber \\ \fl  \left. 
	- \frac{ | \kappa_t |^2  }{1+\nu_t}  +  \frac{1}{ \frac{2}{1+ e^{-2r}}  -  \frac{\nu_t}{1+\nu_t}  }  \left( \frac{\mathrm{Re} \kappa_t}{1+ \nu_t}  + \frac{2 \mathrm{Re} \alpha}{ 1+ e^{-2r} } \right)^2 
	+ \frac{1}{ \frac{2}{1+ e^{2r}}  -  \frac{\nu_t}{1+\nu_t}  }  \left( \frac{ - \mathrm{Im} \kappa_t }{1+ \nu_t}  + \frac{2 \mathrm{Im} \alpha}{ 1+ e^{2r} } \right)^2 
	 \right. \nonumber \\ \fl  \left. 
	- \frac{ 2 (\mathrm{Re} \alpha )^2 }{1+ e^{-2r}}   -    \frac{ 2 (\mathrm{Im} \alpha )^2}{1+ e^{2r}}  \right]	.   \label{sqchrfunc2} 
\end{eqnarray}

The Monte Carlo paths for an initial squeezed state is shown in figure~\ref{fig_num1} (c).
We take the initial parameters with $r =1.2$ and $\alpha = 1.$
The change in the average photon number upon photodetection
is negative because of the super-Poissonian photon-number distribution.

\section{Conclusion}

We have discussed the
simultaneous measurement process of photon-counting and homodyne detection, 
and derived the corresponding stochastic Schr\"odinger equation.
This stochastic equation describes the time evolution of the quantum state 
under a given sequence of measurement outcomes.
The analytical expression of the conditional wave function is obtained and, using this expression, we have derived 
the probability density function and the generating functional of measurement records as a functional of the initial state of the system.
We have also derived the expectation values of the homodyne records conditioned on a photon-counting event.
These analytic results on the cross-correlations between
two measurement outputs quantitatively show
the wave-particle duality of the radiation in quantum measurement.
That is, the measurement backaction of one-type of measurement influences the measurement outcomes of the other type, and vice versa.
We have applied these general results to four typical initial conditions:
coherent, number, thermal, and squeezed states.
For each of these initial states,
we have obtained analytic expressions 
of the generating functional of the measurement records
and showed that the nontrivial correlations between two output channels originating from measurement backaction.
We have performed Monte Carlo simulations of the average photon number,
which show the combined nature of photon-counting and homodyne detection,
implying the particle-wave duality of the photon field.

\ack
This work was supported by
KAKENHI 22340114, a Grant-in-Aid for Scientiﬁc Research on Innovation Areas
“Topological Quantum Phenomena” (KAKENHI 22103005), a Global COE Program
“the Physical Sciences Frontier”, and the Photon Frontier Network Program, from MEXT of Japan. 
Y. K. acknowledges the ALPS for financial support.

\appendix

\section{Derivation of equation (\ref{sqchar2})}
In this appendix, we derive equation (\ref{sqchar2}) based on the Q-function technique.

The Q-function~\cite{gardiner_zoller} of state $\hrho$ is defined by
\begin{equation}
	Q(\beta , \beta^*) := \frac{1}{\pi} \bra{\beta}  \hrho \ket{\beta}.  \label{defqfunction}
\end{equation}
This function is convenient for calculating the expectation values of antinormally ordered operators.
Let $f(\beta,\beta^*)$ be an arbitrary function of complex variables $\beta, \beta^*$ 
with an expression
\begin{eqnarray}
	f(\beta ,\beta^*) = \sum_{r,s} f_{rs} \beta^r (\beta^*)^s.
\end{eqnarray}
The antinormally ordered operator of $f(\beta ,\beta^*)$ is defined by
\begin{eqnarray}
	\mathcal{A} [  f(\ha, \ha^\dagger) ] :=  \sum_{r,s} f_{rs} \ha^r (\ha^{\dagger })^s  .
\end{eqnarray}
Then, from the overcompleteness relation of coherent states, we have~\cite{gardiner_zoller}
\begin{eqnarray}
	\mathrm{tr} \left(  \hrho \mathcal{A} [  f(\ha, \ha^\dagger) ]  \right)
	&=   \sum_{r,s} f_{rs}   \mathrm{tr} \left( \hrho \ha^r   \int \frac{ d^2 \beta }{\pi}  \ket{\beta } \bra{\beta }   (\ha^{\dagger })^s   \right)  \nonumber \\
	&= \int \frac{ d^2 \beta }{\pi}  \sum_{r,s} f_{rs}  \beta^r (\beta^{*})^s   \mathrm{tr} \left( \hrho     \ket{\beta } \bra{\beta }     \right)  \nonumber \\
	&= \int d^2 \beta  f(\beta , \beta^*) Q(\beta , \beta^*) . \label{qfuncex}
\end{eqnarray}
Equation (\ref{qfuncex}) implies that the Q-function can be interpreted as a quasi-probability distribution function for antinormally ordered operators.
To exploit this property, we will calculate the Q-function of the squeezed state $\ket{\alpha , r}$ and 
the antinormally ordered expression of the normally-ordered operator $: \exp [ \kappa \ha + \kappa^* \ha^\dagger + \nu \ha^\dagger \ha   ] :$.

To evaluate the Q-function of the squeezed state $\ket{\alpha , r}$,
we first prove the following formula:
\begin{eqnarray}
	S(r ) = \sqrt{ \frac{1}{\cosh r}  } e^{- \frac{ (\ha^\dagger)^2 }{2} \tanh r   }  e^{ - \ha^\dagger \ha \ln \cosh r }  e^{ \frac{\ha^2}{2}  \tanh r } .  \label{ukousiki}
\end{eqnarray}
To show this, we differentiate the rhs with respect to $r$:
\begin{eqnarray}
	\fl \frac{d}{dr} (\mathrm{rhs}) 
	\nonumber \\
	\fl =  \sqrt{\frac{1}{\cosh r}}  \left[  - \frac{\tanh r}{2} e^{- \frac{ ( \ha^\dagger)^2 }{2} \tanh r   }  e^{ - \ha^\dagger \ha \ln \cosh r }  e^{ \frac{\ha^2}{2}  \tanh r }  
	\right.  \nonumber \\   \left. 
	- \frac{ (\ha^\dagger)^2 }{2 \cosh^2 r}  e^{- \frac{ (\ha^\dagger)^2 }{2} \tanh r   }  e^{ - \ha^\dagger \ha \ln \cosh r }  e^{ \frac{\ha^2}{2}  \tanh r }  
	\right.  \nonumber \\   \left. 
	+ e^{- \frac{ (\ha^\dagger)^2 }{2} \tanh r   } ( - \ha^\dagger \ha \tanh r )  e^{ - \ha^\dagger \ha \ln \cosh r }  e^{ \frac{\ha^2}{2}  \tanh r } 
	\right.  \nonumber \\  \left. 
	+e^{- \frac{ (\ha^\dagger)^2 }{2} \tanh r   }  e^{ - \ha^\dagger \ha \ln \cosh r }  \left(  \frac{\ha^2}{2 \cosh^2 r}  \right)   e^{ \frac{\ha^2}{2}  \tanh r }    \right]  \nonumber \\
	\fl =  \frac{ \ha^2 - (\ha^\dagger)^2 }{2}  (\mathrm{rhs})  \label{appdifeq1}
\end{eqnarray}
In deriving the last equality, we used the following relations:
\begin{eqnarray}
	e^{-\gamma (\ha^\dagger)^2} \ha e^{ \gamma (\ha^\dagger )^2} &= \ha + 2\gamma \ha^\dagger, \\
	e^{-\gamma (\ha^\dagger)^2}  e^{- \lambda \ha^\dagger \ha } \ha  e^{ \lambda \ha^\dagger \ha } e^{\gamma (\ha^\dagger)^2}   &= e^\lambda (\ha + 2\gamma \ha^\dagger). 
\end{eqnarray}
Equation (\ref{appdifeq1}) shows that the rhs satisfies the same differential equation for the lhs.
Since $(\mathrm{lhs}) = (\mathrm{rhs})  = I$ when $r=0$, (\ref{ukousiki}) holds for arbitrary $r$.

Using (\ref{ukousiki}) and 
\[
	   D(-\beta) D (\alpha) = e^{\frac{1}{2}(\alpha \beta^* - \alpha^* \beta) } D(\alpha - \beta ) ,
\]
we obtain
\begin{eqnarray*}
	\bra{\beta } D(\alpha ) S(r) \ket{0} &= \bra{0} e^{\frac{1}{2}(\alpha \beta^* - \alpha^* \beta ) } D(\alpha - \beta ) 
	\sqrt{ \frac{1}{\cosh r}  } e^{- \frac{ (\ha^\dagger)^2 }{2} \tanh r   }  e^{ - \ha^\dagger \ha \ln \cosh r }  e^{ \frac{\ha^2}{2}  \tanh r }  \ket{0}  \\
	&= \frac{ e^{\frac{1}{2}(\alpha \beta^* - \alpha^* \beta ) } }{\sqrt{\cosh r}} \bra{\beta - \alpha}  e^{- \frac{ (\ha^\dagger)^2 }{2} \tanh r   } \ket{0 }\\
	&= \frac{ e^{\frac{1}{2}(\alpha \beta^* - \alpha^* \beta ) } }{\sqrt{\cosh r}} e^{ -\frac{1}{2} (\beta^* - \alpha^*)^2 \tanh r  - \frac{1}{2} | \beta -\alpha |^2   }.
\end{eqnarray*}
Therefore, the Q-function for the squeezed state is given by
\begin{eqnarray} 
	&Q(\beta, \beta^*) \nonumber 
	=  \frac{1}{\pi}  | \bra{\beta} D(\alpha) S(r) \ket{0} |^2 \nonumber \\
	&= \frac{1}{\pi  \cosh r} \exp \left[  - | \beta - \alpha |^2 - \frac{\tanh r}{2} \left\{   (\beta - \alpha)^2 + (\beta^*  - \alpha^*)^2   \right\}   \right] .
\end{eqnarray}

Next, we derive the antinormally-ordered expression of the operator
\begin{eqnarray}
	: \exp [ \kappa \ha + \kappa^* \ha^\dagger + \nu \ha^\dagger \ha   ] :  = e^{ \kappa^* \ha^\dagger  } \left(  : e^{\nu \ha^\dagger \ha} :  \right)  e^{\kappa \ha } .  \nonumber
\end{eqnarray}
From 
\begin{eqnarray}
	\mathcal{A}[  (\ha \ha^\dagger)^m ] & = (\hn + m )(\hn +m-1) \cdots (\hn +1), \\
	 \mathcal{A} [  e^{ x \ha \ha^\dagger  } ] & = \sum_{m=0}^\infty \frac{x^m}{m!} (\hn + m )(\hn +m-1) \cdots (\hn +1) \nonumber \\
	 &= (1-x)^{- \hn -1},
\end{eqnarray}
we have
\begin{eqnarray}
	 :   e^{\nu \ha^\dagger \ha }     :  &=  (1+\nu)^{ \ha^\dagger \ha } \nonumber \\
	&= (1+ \nu)^{-1} \left( 1 - \frac{ \nu }{1 + \nu}  \right)^{  -\ha^\dagger \ha -1  } \nonumber \\
	&=(1+ \nu)^{-1}  \mathcal{A} [  e^{ \frac{\nu}{1+ \nu} \ha \ha^\dagger }] .
\end{eqnarray}
Using
\begin{eqnarray}
	e^{- \kappa \ha} \ha^\dagger e^{ \kappa \ha } &=  \ha^\dagger - \kappa, \\
	e^{ \kappa^* \ha^\dagger} \ha e^{- \kappa^* \ha^\dagger } &=  \ha - \kappa^*,
\end{eqnarray}
we obtain
\begin{eqnarray}
	&: \exp [ \kappa \ha + \kappa^* \ha^\dagger + \nu \ha^\dagger \ha   ] :  \nonumber \\
	&=  (1+ \nu)^{-1} e^{\kappa^* \ha^\dagger} \mathcal{A} [  e^{ \frac{\nu}{1+ \nu} \ha \ha^\dagger }] e^{\kappa \ha} \nonumber \\
	& = \frac{ e^{- |\kappa|^2} }{ 1+ \nu }   e^{\kappa \ha} \mathcal{A} [  e^{ \frac{\nu}{1+ \nu} (\ha - \kappa^*) (\ha^\dagger - \kappa) }]
	e^{\kappa^* \ha^\dagger}  \nonumber  \\
	&= (1+ \nu)^{-1}  \mathcal{A} \left[ \exp \left[  \frac{1}{1+ \nu} \left( - |\kappa|^2 + \kappa \ha + \kappa^* \ha^\dagger + \nu \ha \ha^\dagger  \right)  \right]  \right]    \nonumber \\
	&= \frac{e^{-  \frac{ |\kappa|^2}{ 1+ \nu }  }}{(1+ \nu)} \mathcal{A} [ \exp [ \kappa^\prime \ha  + \kappa^{\prime *} \ha^\dagger + \nu^\prime \ha \ha^\dagger ]  ], \label{antinormalexpression}
\end{eqnarray}
where
\begin{eqnarray*}
	\kappa^\prime &:= \frac{\kappa}{1+\nu }, \\
	\nu^\prime &:= \frac{\nu}{1+ \nu}.
\end{eqnarray*}

We can then evaluate Eq. (\ref{sqchar2}).
The relevant part is
\begin{eqnarray}
	 \fl \bra{\alpha ,r} \mathcal{A} [ \exp [ \kappa^\prime \ha  + \kappa^{\prime *} \ha^\dagger + \nu^\prime \ha \ha^\dagger ]  ] \ket{\alpha ,r} \nonumber \\
	\fl = \int d^2 \beta  Q(\beta , \beta^*) \exp [\kappa^\prime \beta  + \kappa^{\prime *} \beta^* + \nu^\prime |\beta |^2 ] \nonumber \\
	\fl =  
	\left[  \left(1-  \frac{ \nu^\prime (1- e^{-2r}) }{2} \right)   
		   \left(1-  \frac{ \nu^\prime (1- e^{2r}) }{2}   \right)   
	\right]^{-1/2} 
	\nonumber \\ \fl  
	\quad \times
	\exp \left[ \frac{ 1 }{ \frac{2}{1+ e^{-2r}}  - \nu^\prime  }  \left(   \mathrm{Re} \kappa^\prime  + \frac{ 2 \mathrm{Re} \alpha }{1+ e^{-2r}}  \right)  
	+\frac{ 1 }{ \frac{2}{1+ e^{2r}}  - \nu^\prime  }  \left(  - \mathrm{Im} \kappa^\prime  + \frac{ 2 \mathrm{Im} \alpha }{1+ e^{2r}}  \right)
	\right. \nonumber \\   \left.
	- \frac{ 2 ( \mathrm{Re} \alpha )^2 }{1+ e^{-2r}}
	- \frac{ 2 ( \mathrm{Im} \alpha )^2 }{1+ e^{2r}}
	\right]  .
\end{eqnarray}
Going back to original $\kappa$ and $\nu$, we obtain equation (\ref{sqchar2}).


\section*{References}
\begin{thebibliography}{99}
\bibitem{neumann}
von Neumann J 1932
\textit{Mathematische Grundlagen der Quantenmechanik,}
(Berlin, Springer);
English translation \textit{Mathematical Foundations of Quantum Mechanics}, transl. Robert T. Beyer, (Princeton, Princeton University Press, 1955).
\bibitem{davies_lewis}
Davies E B and Lewis J T 1970 \textit{Commun. math. Phys.} \textbf{17} 239
\bibitem{davies_book}
Davies E B, 1976 \textit{Quantum Theory of Open Systems} (New York: Academic Press)
\bibitem{kraus_wootters}
Kraus K, B\"ohm A, Dollard J T and Wootters J T 1983
\textit{States, Effects, and Operations: Fundamental Notions of Quantum Theory,} Lecture Notes in Physics, Vol. 190 (Berlin: Springer)
\bibitem{lindblad} 
Lindblad G 1976 \textit{Commun. Math. Phys.} \textbf{48} 119
\bibitem{gorini_kossakowski_sudarshan}
Gorini V, Kossakowski A and Sudarshan E C G
1976 \textit{J. Math. Phys.} \textbf{17} 821
\bibitem{jacobs_steck1}
Jacobs K and Steck D A
2006
\textit{Contemp. Phys.} \textbf{47} 279
\bibitem{davies_qsp}
Davies E B 1969 \textit{Commun. Math. Phys.} \textbf{15} 277;
1970 \textit{Commun. Math. Phys.} \textbf{19} 83;
1971 \textit{Commun. Math. Phys.} \textbf{22} 51
\bibitem{diosi} 
Di\'osi L
1988
\PL A \textbf{129} 419
\bibitem{wiseman_diosi}
Wiseman H M and Di\'osi L 2001 \textit{Chem. Phys.} \textbf{268} 91
\bibitem{barchielli_paganoni_zucca}
Barchielli A, Paganoni A M and Zucca F
1998
\textit{Stochastic Process. Appl.}
\textbf{73} 69
\bibitem{zurek}
Zurek W H 1981 \PR D \textbf{24} 1516;
1982 \PR D \textbf{26} 1862
\bibitem{presilla}
Presilla C, Onofrio R and Tambini U 1996 \textit{Ann. Phys. (NY)}
\textbf{248} 95
\bibitem{dalibard}
Dalibard J, Castin Y and M\/olmer K 1992 \PRL \textbf{68} 580
\bibitem{gisin}
Gisin N and Percival I 1992 \JP A \textbf{25} 5677;
1992 \PL A \textbf{167} 315
\bibitem{breuer_petruccione}
Breuer H P and Petruccione F 2002 \textit{The Theory of Open Quantum Systems} (Oxford University Press)
\bibitem{haroche_raimond}
Haroche S and Raimond J M 2006 \textit{Exploring the Quantum: Atoms, Cavities and Photons}
(Oxford University Press)
\bibitem{mensky}
Mensky M B 1979 \PR D \textbf{20} 384;
1979 \textit{Zh. Eksp. Teor. Fiz. } \textbf{77} 1326
(English translation: 1980 \textit{Sov. Phys. JETP} \textbf{50}(4) 667)
\bibitem{chantasri}
Chantasri A, Dressel J and Jordan A N 2013 arXiv:1305.5201
\bibitem{srinivas_davies} 
Srinivas M D and Davies E B 1981 \textit{Opt. Acta} \textbf{28} 98;
Mandel L, 1981 \textit{Opt. Acta} \textbf{28} 1447;
Srinivas M D and Davies E B 1981 \textit{Opt. Acta} \textbf{29} 235 
\bibitem{ueda}
Ueda M 1989 \textit{Quantum Opt.} \textbf{1} 131
\bibitem{ueda_ogawa}  
Ueda M, Imoto N and Ogawa T  
1990
\PR A \textbf{41} 3891 
\bibitem{imoto_ueda_ogawa}
Imoto N, Ueda M and Ogawa T
1990 
\PR A \textbf{41} 4127
\bibitem{haroche_qnd}
Gleyzes S, Kuhr S, Guerlin C, Bernu J, Del\'eglise S, Hoff U B,
Brune M, Raimond J M and  Haroche S 2007 {\it Nature} {\bf 446} 297
\bibitem{barchielli_book}
Barchielli A and Gregoratti M 2009 \textit{Quantum Trajectories
and Measurements in Continuous Time: The Diffusive Case},
Lecture Notes in Physics, 782 (Springer)
\bibitem{carmichael1}  
Carmichael H J
1993
\textit{An Open Systems Approach to Quantum Optics},
Lecture Notes in Physics, New Series m: Monographs,
m18 (Berlin: Springer)
\bibitem{wiseman_milburn1} 
Wiseman H M and Milburn G J
1993
\PR A \textbf{47} 642
\bibitem{barchielli_holevo}
Barchielli A and Holevo A S 1995 \textit{Stochastic Processes and their Applications} \textbf{58} 293
\bibitem{pellegrini}
Pellegrini C 2007 \textit{Annales de l'Institut Henri Poincar\'e - Probabilit\'es et Statistiques} \textbf{46} 924
\bibitem{carmichael_orozco}
Carmichael H J, Castro-Beltran H M, Foster G T and Orozco L A
2000
\PRL \textbf{85} 1855
\bibitem{wiseman}
Wiseman H M
2002
\PR A \textbf{65} 032111
\bibitem{viola_onofrio}
Viola L and Onofrio R 1997 \PR A \textbf{55} 3291
\bibitem{jacobs_steck2}
Jacobs K and Steck D A
2011
\NJP \textbf{13} 013016
\bibitem{gardiner1}
Gardiner C W
2009
\textit{Stochastic Methods  A Handbook for the Natural and Social Sciences} 4th edition,
(Berlin: Springer)
\bibitem{jacobs_book}
Jacobs K
2010
\textit{Stochastic Processes for Physicists: Understanding Noisy Systems}
(Cambridge University Press)
\bibitem{wiseman_milburn2} 
Wiseman H M and Milburn G J
2010
\textit{Quantum Measurement and Control},
(Cambridge University Press)
\bibitem{gardiner_zoller}
Gardiner C W and Zoller P
2004
\textit{Quantum Noise} 3rd edition
(Berlin: Springer)
\end{thebibliography}
\end{document}